\documentclass[12pt, draftclsnofoot, onecolumn]{IEEEtran}

\usepackage[T1]{fontenc}

\usepackage{cite}
\usepackage{graphicx}
\usepackage{amsmath}
\usepackage{algorithmic}
\usepackage{array}
\usepackage{stfloats}
\usepackage{url}
\usepackage[caption=false,font=normalsize,labelfont=sf,textfont=sf]{subfig}

\usepackage{amsfonts}

\newtheorem{definition}{Definition}

\graphicspath{{./figures/}}

\begin{document}
%
\title{User Clustering for Multicast Precoding\\ in Multi-Beam Satellite Systems}


\author{Alessandro~Guidotti,~\IEEEmembership{Member,~IEEE,}
        Alessandro~Vanelli-Coralli,~\IEEEmembership{Senior Member,~IEEE,}
        Giorgio~Taricco,~\IEEEmembership{Senior Member,~IEEE,}
        and~Guido~Montorsi,~\IEEEmembership{Senior Member,~IEEE}
\thanks{A. Guidotti and A. Vanelli-Coralli are with the Department of Electrical, Electronic, and Information Engineering, University of Bologna, 40134, Bologna, Italy (email:\{a.guidotti,alessandro.vanelli\}@unibo.it).}
\thanks{G. Taricco and G. Montorsi are with the Department of Electronics and Telecommunications, Politecnico di Torino, 10129, Torino, Italy (email:\{giorgio.taricco,guido.montorsi\}@polito.it).}
}


\maketitle

\begin{abstract}
Current State-of-the-Art High Throughput Satellite systems provide wide-area connectivity through multi-beam architectures. However, due to the tremendous system throughput requirements that next generation Satellite Communications expect to achieve, traditional 4-colour frequency reuse schemes, \emph{i.e.}, two frequency bands with two orthogonal polarisations, are not sufficient anymore and more aggressive solutions as full frequency reuse are gaining momentum. These approaches require advanced interference management techniques to cope with the significantly increased inter-beam interference, like multicast precoding and Multi User Detection. With respect to the former, several peculiar challenges arise when designed for SatCom systems. In particular, to maximise resource usage while minimising delay and latency, multiple users are multiplexed in the same frame, thus imposing to consider multiple channel matrices when computing the precoding weights. In this paper, we focus on this aspect by re-formulating it as a \emph{clustering problem}. After introducing a novel mathematical framework, we design a k-means-based clustering algorithm to group users into simultaneously precoded and served clusters. Two similarity metrics are used to this aim: the users' Euclidean distance and their channel distance, \emph{i.e.}, distance in the multidimensional channel vector space. Through extensive numerical simulations, we substantiate the proposed algorithms and identify the parameters driving the system performance.
\end{abstract}


\IEEEpeerreviewmaketitle

\IEEEdisplaynontitleabstractindextext

%

\section{Introduction}
\IEEEPARstart{D}{uring} the last years, in order to provide significantly larger throughput to the on-ground users, Satellite Communication (SatCom) systems evolved from the traditional single-beam architecture to a more advanced multi-beam deployment. Current State-of-theArt (SoA) High Throughput Satellite (HTS) systems in Ka-band provide connectivity through $80-100$ beams to cover a single region as, for instance, Europe, \cite{Intro6}. The basic principle behind this type of system architecture and deployment is the frequency reuse concept, which is well-known in the terrestrial community. Today, SoA HTS systems split the available bandwidth into two frequency bands with two orthogonal polarisations, leading to a 4-colour frequency reuse scheme. However, when aiming at providing terabit connectivity through satellite systems, \cite{Intro5}, different architectures and technologies are required mainly to increase the system throughput by either augmenting the available spectrum or the spectral efficiency at system level. Notably, current physical layer (PHY) technologies already bring the link spectral efficiency close to the theoretical limits and, thus, more advanced solutions shall be defined. On the one hand, in the past years, several activities focused on defining more flexible spectrum usage paradigms based on spectrum sharing through Cognitive Radio techniques, \cite{Intro1,Intro2,Intro7}. On the other hand, more aggressive frequency reuse schemes can be targeted in order to fully exploit the available licensed bandwidth and target much larger throughput levels. This is the case of \emph{full frequency reuse} architectures. In this context, the overall system performance is strongly limited by the interference between adjacent beams, due to the side-lobes of the beams radiation patterns. It is, thus, of paramount importance to implement advanced interference mitigation techniques so as to counterbalance the increased inter-beam interference levels introduced by the iso-frequency allocation paradigm, at the expense of an increased system complexity, such as precoding, \cite{Prec1,Prec2,ESAMIMO}, and Multi-User Detection (MUD) at the receiver, \cite{Intro8}.

The success of multi-user Multiple-Input Multiple-Output (MU-MIMO) techniques in terrestrial communications, together with the introduction of the super-frame structure in the DVB-S2X standard, \cite{DVBS2X}, led the Satellite Communications community to assess and implement precoding techniques in multi-beam HTS systems. Building on the early works in \cite{Prec1,Prec2,Prec3}, which paved the way for the application of MU-MIMO in HTS systems, the SatCom community has been intensifying the work on precoding-based satellite systems, as demonstrated by the significant contributions brought to this topic in \cite{Prec4,Prec5,Prec6,Prec7,Prec8,Prec8_1,Prec10,Prec11}. These works showed that significant gains can be obtained in the overall system throughput by means of both linear and non-linear precoding techniques over the multi-beam fixed satellite channel. This channel poses peculiar challenges and has quite different behaviours from the terrestrial one. Since linear precoding techniques provide already significant throughput gains with a limited complexity, they might be preferred with respect to more complex techniques, \emph{i.e.}, non-linear precoding, for which the increase in the system complexity is not justified by an equivalent performance benefit, \cite{Intro9}.

While significant gains can be obtained by means of precoding techniques, it shall be highlighted that they are based on the transmission of independent data to different users. As a matter of facts, SatCom systems provide large single-link data rates to the users and, thus, in order to enhance the overall system level efficiency multiple users shall be multiplexed into the same PHY codeword. Within this framework, traditional single-user (unicast) precoding is not always applicable and multicast precoding techniques can provide an efficient solution as shown in several recent works, \cite{Prec12,Prec13,Prec14,CTTCprec,Taricco}. Within this framework, one of the most critical factors is related to the approach with which users are grouped within the same frame, because it directly and deeply impacts the overall system performance in terms of achievable throughput, as preliminarily analysed in \cite{Taricco,Prec14,Prec8}. We propose a novel design approach to the users grouping for multicast precoding based on \emph{clustering theory} concepts. The definition of this framework allows us to identify and highlight the pivotal aspects that need to be taken into account for the design of a multi-beam HTS system based on multicast precoding.

\subsection{Previous works}
The first implementations of the precoding concept to a Satellite Communication context is provided in \cite{Prec1,Prec2}. In these early works, the authors moved from the linear MU-MIMO techniques already substantiated in terrestrial cellular systems and proposed both Zero-Forcing (ZF) and Minimum Mean Square Error (MMSE) algorithms to enhance the throughput on the Forward Link (FL) of a multi-beam satellite system. These works also provided some considerations on the main implementation challenges that impact the effectiveness of precoding in SatCom systems, together with \cite{Prec3}, in which the impact of partial Channel State Information (CSI) at the transmitter (CSIT) has been addressed. In \cite{Prec11}, the implementation of precoding to the DVB-S2X standard is discussed based on the outcome of several European Space Agency (ESA) R\&D activities. In particular, practical challenges related to the implementation of precoding to HTS systems are discussed, like framing issues, imperfect channel estimation, outdated phase estimates, and multiple gateways. An extensive analysis of practical impairments for the implementation of precoding in DVB-S2X systems is also provided in \cite{Prec6}. In \cite{Prec5}, the authors provide a review of several precoding techniques and propose an optimisation of the linear precoding design. In particular, general linear and non-linear power constraints are addressed by means of an iterative algorithm that optimises the precoding vectors and the power allocations in an alternating fashion. An extensive review of precoding techniques for multi-beam systems is also provided in \cite{Prec7}. The authors of \cite{Prec4} implemented the Tomlinson-Harashima precoding (THP), a non-linear precoding technique based on modulo operations over the constellation symbols, also taking into account the beam gain. In \cite{Prec10}, the authors assess the performance of linear beamforming in terms of satisfying specific traffic demands by including generic linear constraints the transmit covariance matrix. 

The first work providing the design of multicast precoding for satellite systems was based on a regularised channel inversion, \cite{Taricco}. In this work, the author proposed to consider all the users to be served together as a single user with an equivalent channel matrix equal to the average of the single channel matrices. In this work, a geographical user grouping is also proposed in which users to be precoded together are chosen based on their geographical position, under the assumption that the same number of users are precoded together across the different beams. The computation of the precoding matrix is based on the pragmatic approach proposed in \cite{CTTCprec}, in which the authors jointly design the linear precoding and ground-based beamforming at the gateway. An alternative solution to the computation of the precoding matrix is provided in \cite{Prec8_1}, in which the authors implement a technique based on block Singular Value Decomposition (SVD). In this case, the precoding matrix is built row-wise and the performance is actually improved, although with a significant increase in the system computational cost. The authors of \cite{Prec8} discuss on the implementation of linear precoding techniques to multi-beam broadband fixed satellite communications by also also introducing some preliminary consideration on the issues related to users grouping. In \cite{Prec12}, the optimisation problem of multicast precoding with per-antenna power constraints is addressed, for which the authors propose an optimal, although computationally costly, solution. In \cite{Prec13}, the authors focus on framing multiple users per transmission and on the per-antenna transmit power limitations and propose a solution for frame-based precoding based on the principles of physical layer multicasting to multiple co-channel groups under per-antenna constraints. Finally, in \cite{Prec14}, a two stage linear precoding is proposed to lower the complexity in the ground segment, under the presence of non-ideal CSI as well. In addition, some aspects related to user grouping are also discussed, in particular referring to a grouping based on the channel coefficients.

\subsection{Contributions and Paper Organisation}

In this work, we focus on multicast precoding techniques for multi-beam satellite systems and, in particular, on the choice of the users to be grouped together within the same frame. As shown in the following, this problem can be formulated as a \emph{clustering problem}, for which different algorithms and approaches have been discussed in the literature. 

Aspects related to the user grouping were mainly introduced in \cite{Taricco,Prec14}. However, in these works there are two assumptions that might lead to sub-optimal performance: i) the number of users to be grouped into the same frame is assumed to be constant across the beams; and ii) at each time frame, a randomly chosen user is selected as reference to define the remaining group members. The former aspect might introduce non-ideal groups into the system, since, depending on the density with which users are distributed, in some beams there might be dense enough areas to have the targeted group cardinality with a limited distance among users, while in others they might be far away from each other. Since in multicast precoding the performance is driven by the user with the lowest SINR, this assumption might result in a quite limited performance. The latter point highlighted above can introduce a sub-optimal grouping for similar reasons: in case the selected user is a so called \emph{outlier} in the system, \emph{i.e.}, it is located far away from the other users, then the performance will be significantly degraded. It shall be noted that channel vectors are used as metrics in \cite{Prec14}, instead of geographical distances as in \cite{Taricco}. However, in terms of user grouping, the above issues are the same, as it will be explained in the following. In addition, in \cite{Taricco,Prec14} the proposed scheduling assumed that, during each time frame, the same number of users to be served in all of the beams, denoted by $Q$, is chosen by randomly selecting one user and then identifying the $Q-1$ closest users, depending on the similarity metric. During the following time frame, there is another user drop over the beams coverage and a successive random selection of the first reference user, which implies that not all of the users initially dropped over the coverage area, which can be a number as high as several hundreds depending on the user density, are served, but only $Q$ per beam, thus leading to a performance upper bound.

Moving from these works, and from previous multicast precoding implementations to SatCom in which the user grouping was not the pivotal aspect, in this paper we provide the following contributions:
\begin{itemize}
    \item we consider a variable number of users per beam and we also assume a variable number of users per group to be precoded within the same frame;
    \item we provide a mathematical framework for the design of multicast precoding systems based on clustering theory and algorithms. In this context, details on the initial selection of the users so as to avoid the sensitivity to outliers are provided;
    \item we design a k-means based clustering algorithmaimed at providing minimum variance partitions of the initial users pool. In particular, two metrics will be considered: i) the Euclidean distance; and ii) the distance in the multi-dimensional space of user channel coefficients;
    \item we provide an extensive and thorough numerical simulation campaign so as to identify the main limiting factors and benefits in implementing proper clustering algorithms for HTS based on multicast precoding.
\end{itemize}

The remainder of this paper is organised as follows. In Section~\ref{sec:SystemModel}, we introduce the system model in terms of system architecture, multicast precoding algorithm, and problem statement. In Section~\ref{sec:Clustering}, we formulate the users grouping problem of multicast precoding in terms of a clustering problem based on the \emph{k-means} clustering algorithm, \cite{kmeansPP}, also proposing two different metrics to be optimised by the clustering algorithm. In Section~\ref{sec:Results}, we provide the numerical results of the extensive simulation campaign in which the average rate per beam is obtained as a function of the number of users precoded in the same frame and of the distance among them. Finally, Section~\ref{sec:Conclusion} concludes this paper.
\paragraph*{Notation}
Throughout this paper, and if not otherwise specified, the following notation is used: bold face lower case and bold face upper case characters denote column vectors and matrices, respectively. ${(\cdot)}^T$ denotes the matrix transposition operator. ${(\cdot)}^H$ denotes the matrix conjugate transposition operator. $\parallel\cdot\parallel$ denotes the Euclidean norm. $\mathbb{E}\left\{\cdot\right\}$ denotes the expectation operator. $\mathrm{diag}(\cdot)$ denotes a diagonal matrix. $\mathbf{I}_N$ denotes the $N\times N$ identity matrix. $\mathcal{N}_c\left(\mu,\sigma^2\right)$ denotes the circularly-symmetric complex Gaussian distribution with mean $\mu$ and variance $\sigma^2$. $\left|\mathcal{A}\right|$ denotes the cardinality of set $\mathcal{A}$.

\section{System Model and Problem Statement}
\label{sec:SystemModel}
\begin{figure}[!t]
\centering
\includegraphics[width=0.55\columnwidth]{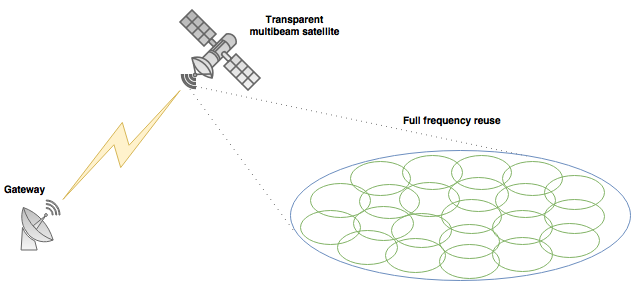}
\caption{System architecture.}
\label{fig:Architecture}
\end{figure}

We consider a Geostationary Earth Orbit (GEO) HTS system operating with full frequency reuse to provide broadband connectivity by means of multiple beams, as shown in Fig.~\ref{fig:Architecture}. The following general assumptions are made with respect to the system architecture and operations: i) the satellite is assumed to be transparent; ii) linear precoding techniques are implemented on the Forward Link (FL), which can then be modelled as a Multiple Input Multiple Output (MIMO) channel; iii) the satellite payload is equipped with $N_B$ antennas and each antenna radiates its signal to the corresponding on-ground beam, \emph{i.e.}, we consider $N_B$ on-ground beams and the FL channel is thus represented by a $N_B\times N_B$ matrix; and iv) a single gateway (GW) manages the Channel State Information (CSI), assumed to be ideal, obtained from the Return Link (RL) in order to compute the precoding weights.

\subsection{System Model}
\subsubsection{Linear Precoding}
We assume that Time Division Multiple Access (TDMA) is implemented to serve the users in the $N_B$ beams in each time frame. Thus, during each time frame, the GW relies on the CSIT previously obtained from the RL and simultaneously serves $N_B$ users by implementing linear precoding algorithms.\\
In this paper, we consider the work in \cite{Taricco,CTTCprec} that propose a pragmatic linear precoding approach based on the duality principle, \cite{DualityPaper}. In particular, while the aim of \cite{DualityPaper} was originally to characterise the sum capacity of the Gaussian Broadcast channel, in \cite{CTTCprec} the authors showed that it can be elegantly used to derive a Minimum Mean Square Error (MMSE) precoder.

In the system context outlined in the previous section, and in the presence of Additive White Gaussian Noise (AWGN), the received signal can be modelled by the following equation over a linear channel:
\begin{equation}
\label{eq:RXsignal}
    \mathbf{y} = \mathbf{H}\hat{\mathbf{x}} + \mathbf{z}
\end{equation}
where: i) $\mathbf{y}$ is the $N_B\times 1$ vector of received symbols; ii) $\mathbf{z}$ is a $N_B\times 1$ vector of independent and identically distributed (i.i.d.) circularly-symmetric complex Gaussian random variables with zero mean and unit variance, \emph{i.e.}, $z_i\sim\mathcal{N}_c\left(0,1\right)$, $i=1,\ldots,N_B$; iii) $\mathbf{H}\in\mathbb{C}^{N_B\times N_B}$ is a $N_B\times N_B$ complex channel matrix, in which $h_{ij}$ represents the channel gain between the $i$-th user and the $j$-th antenna feed on the satellite, with $i,j=1,\ldots,N_B$; and iv) $\hat{\mathbf{x}}$ is the precoded signal transmitted from the satellite antenna feeds, \emph{i.e.}:
\begin{equation}
\label{eq:PrecTXsignal}
    \hat{\mathbf{x}} = \mathbf{W}\mathbf{x}
\end{equation}
where $\mathbf{x}={\left(x_1,\ldots,x_{N_B}\right)}^T$ is $N_B\times 1$ vector of user symbols and $\mathbf{W}$ is the $N_B\times N_B$ precoding matrix. Resorting to the uplink-downlink duality principle in \cite{DualityPaper}, the authors of \cite{Taricco,CTTCprec} derived the following optimum MMSE precoding matrix:
\begin{equation}
\label{eq:MMSEprec}
    \mathbf{W} = {\left( \mathbf{H}^H\mathrm{diag}(\mathbf{q})\mathbf{H} + \mathbf{I}_{N_B} \right)}^{-1}\mathbf{H}^H\mathrm{diag}(\mathbf{q})
\end{equation}
where $\mathbf{q}$ is a virtual uplink power vector corresponding to the solution of a non-linear equation system and $\mathbf{I}_{N_B}$ is the $N_B\times N_B$ identity matrix. As highlighted in \cite{Taricco}, one of the main challenges raised by this approach is related to the fact that the precoding matrix modifies the transmitted signal powers. To circumvent this issue, a simple sub-optimal pragmatic approach was proposed in \cite{CTTCprec}, which is based on setting equal transmitted power levels at the satellite antenna feeds and on two normalisation steps: i) first, each column of the precoding matrix is normalised to have unit-norm; and ii) each row of the precoding matrix is further re-scaled to have unit norm. The second normalisation actually loosens the optimality of the precoding algorithm, but since the transmitting power levels were set to be equal across the antenna feeds, this also results in the satisfaction of a per-antenna power constraint, \cite{Taricco}.

\subsubsection{Multicast Precoding}

The above approach provides good performance when implemented for unicast precoding, \emph{i.e.}, when considering a single user per beam. However, as highlighted in \cite{Taricco}, when we must serve multiple users per frame, \emph{i.e.}, for multicast precoding, during the same time frame the MIMO system is described by many channel matrices, one per user, which makes the implementation of linear precoding non-trivial. Assuming that in each beam there are $N_U$ users to be multiplexed into the same DVB-S2X codeword, and in clear-sky propagation conditions, their channel gains depend exclusively on their location and the antenna radiation pattern and can be arranged as row vectors as follows:
\begin{equation}
\label{eq:UserChannelVector}
    \mathbf{h}_b^{(i)}(\mathbf{v}_{b,i}) = \left(h_{b,1}^{(i)}(\mathbf{v}_{b,i}),\ldots,h_{N_B,N_B}^{(i)}(\mathbf{v}_{b,i})\right),\ i=1,\ldots,N_U
\end{equation}
where $\mathbf{v}_{b,i}$ denotes the location vector of the $i$-th user in the $b$-th beam and $h_{b,\ell}^{(i)}(\mathbf{v}_{b,i})$ denotes the channel gain between it and the $\ell$-th satellite antenna feed, $\ell=1,\ldots,N_B$. Since all beams have the same number of users $N_U$, the received signal when considering the $i$-th users in all beams is given by:
\begin{equation}
\label{eq:RXsignal2}
    \mathbf{y}^{(i)} = \mathbf{H}^{(i)}\hat{\mathbf{x}} + \mathbf{z}^{(i)},\ i=1,\ldots,N_U
\end{equation}
where now we have $N_U$ matrix equation describing our system and:
\begin{equation}
    \mathbf{H}^{(i)} = {\left( \mathbf{h}_1^{(i)}(\mathbf{v}_{1,i})^T,\ldots,\mathbf{h}_{N_B}^{(i)}(\mathbf{v}_{N_B,i})^T \right)}^T
\end{equation}
is the system channel matrix when considering the $i$-th user in all beams. In order to provide a single channel matrix to the pragmatic precoding algorithm, in \cite{Taricco} the author proposed a solution based on average channel matrices. Since the $N_B\times N_B$ channel matrix $\mathbf{H}$ required to compute the precoding matrix in (\ref{eq:MMSEprec}) is the combination of several $1\times N_B$ channel gain row vectors, exactly $N_B$, one for each receiver, then we can compute a representative channel vector for each beam as a given function of the different $N_U$ channel vectors. In particular, we can compute an equivalent channel vector for the $b$-th beam:
\begin{equation}
\label{eq:EquivVectorH}
    \tilde{\mathbf{h}}_b = \mathbf{\Xi}\left(\mathbf{h}^{(1)}_b(\mathbf{v}_{b,1}),\ldots,\mathbf{h}^{(N_U)}_b(\mathbf{v}_{b,N_U})\right), \ b=1,\ldots,N_B
\end{equation}
which is now depending only on the selected beam, and not on the specific users to be served within its coverage. Once the above vectors have been defined, they are combined to form the $N_B\times N_B$ equivalent channel matrix:
\begin{equation}
\label{eq:EquivChannelMatrix}
    \mathbf{H}_{\mathbf{\Xi}}={\left( \tilde{\mathbf{h}}_1^T,\ldots,\tilde{\mathbf{h}}_{N_B}^T \right)}^T
\end{equation}
Finally, this matrix is used to compute the precoding matrix as in (\ref{eq:MMSEprec});
\begin{equation}
\label{eq:EquivPrecodingMatrix}
    \mathbf{W}_{\mathbf{\Xi}} = {\left( \mathbf{H}_{\mathbf{\Xi}}^H\mathrm{diag}(\mathbf{q})\mathbf{H}_{\mathbf{\Xi}} + \mathbf{I}_{N_B} \right)}^{-1}\mathbf{H}_{\mathbf{\Xi}}^H\mathrm{diag}(\mathbf{q})
\end{equation}
By observing equations (\ref{eq:EquivVectorH})-(\ref{eq:EquivPrecodingMatrix}), it can be noticed that the system performance is driven by two critical design factors: i) how to select the users to be grouped together for each beam; and ii) the definition of the combination function $\mathbf{\Xi}$. In this paper, we will assume a simple arithmetic mean as combination function, as proposed in \cite{Taricco}, and focus on the users grouping approach, which can be modelled as a \emph{cluster analysis problem}: we must group a set of objects (users) such that objects in the same group (called a cluster) are more similar (in some sense or another) to each other than to those in other groups (clusters). It is worthwhile highlighting that the arithmetic mean is computed based on the amplitudes of the equivalent channel matrix, while the phase contributions are neglected and assumed to be dealt with at the receiver.  

\subsubsection{Channel model}
As outlined in the previous paragraphs, the $N_B\times N_B$ channel matrix $\mathbf{H}^{(i)}$ models the channel between the multi-beam satellite antenna feeds and the $i$-th users in on-ground beams. In particular, the $b$-th row of this matrix, $\mathbf{h}_b^{(i)}$ with $b=1,\ldots,N_B$ and $i=1,\ldots,N_U$, describes the channel gain between the $N_B$ antenna feeds towards the $i$-th user in the $b$-th beam and its elements are defined as follows:
\begin{equation}
\label{eq:ChannelGain}
    h_{bj}^{(i)} = \frac{\sqrt{G_RG_{loss}G_{bj}^{(i)}}}{4\pi \frac{d_i}{\lambda}\sqrt{\kappa T_b B_w}}e^{-j\frac{2\pi}{\lambda}d_i}
\end{equation}
where $G_R$ is the receiver antenna gain, $G_{loss}$ models the overall antenna losses, $G_{bj}^{(i)}$ is the multi-beam antenna gain between the $j$-th antenna feed and the $b$-th receiving beam, $d_i$ is the distance between the GEO satellite and the considered $i$-th user in the $b$-th beam, $\kappa$ is the Boltzmann constant, $T_b$ is the clear-sky noise temperature of the receiving antenna in the $b$-th beam, $B_w$ is the user link bandwidth, and $\lambda$ the carrier wavelength. As previously highlighted, when the equivalent channel matrix in $\mathbf{H}_{\mathbf{\Xi}}$ (\ref{eq:EquivChannelMatrix}) is computed, the phase contribution in (\ref{eq:ChannelGain}) is neglected and assumed to be taken care of at the receiver, \emph{i.e.}, the equivalent precoding matrix $\mathbf{W}_{\mathbf{\Xi}}$ in (\ref{eq:EquivPrecodingMatrix}) is built accounting for the channel amplitude only.

\subsection{Problem Statement}
\label{sec:ProblemStatement}
In the considered framework, we aim at defining a proper clustering strategy for the users such that the performance of multicast precoding is optimal. In particular, we look for a partition of the users served by each beam such that:
\begin{enumerate}
    \item a user exclusively belongs to a single cluster;
    \item the intersection between two different clusters is empty;
    \item the union set of all clusters provides the set of users initially deployed in the beam.
\end{enumerate}
While the first two conditions guarantee that each user is served in one time frame only, as long as not all of the users have been served in the beam, the third one makes us sure that all of the users within a beam are served.
We consider two different aspects: i) the choice of the clustering algorithm; and ii) the choice of the parameter(s) to be optimised by the chosen algorithm, \emph{i.e.}, the similarity metric to measure the clustering performance. These two concepts will be addressed in Section~\ref{sec:Clustering}.

The two above aspects directly and deeply affect the overall system performance in terms of achievable rate. As a matter of fact, in each time frame and in each beam, multiple users with a different Signal-to-Interference plus Noise-Ratio (SINR) are precoded within the same DVB-S2X codeword. By focusing on the $b$-th beam, the signal by the $i$-th user is given by:
\begin{equation}
\label{eq:RXsignal2_b}
    y_b^{(i)} = \mathbf{h}_b^{(i)}\mathbf{w}_{\mathbf{\Xi},b}x_b + \sum_{\ell\neq b}\mathbf{h}_b^{(i)}\mathbf{w}_{\mathbf{\Xi},\ell}x_{\ell} + z_b^{(i)}
\end{equation}
where $\mathbf{h}_b^{(i)}$ denotes the $b$-th row of the $\mathbf{H}^{(i)}$ channel matrix and $\mathbf{w}_{\mathbf{\Xi},b}$ denotes the $b$-th column of the equivalent precoding matrix $\mathbf{W}_{\mathbf{\Xi}}$. Under the assumption of Gaussian inputs and remembering that $z_b^{(i)}\sim\mathcal{N}_c\left(0,1\right)$, the achievable rate for the $i$-th user within the $b$-th beam is given by:
\begin{equation}
\label{eq:SINR_1}
    \mu_i^{(b)} = \log_2\left(1+\gamma_i^{(b)}\right)
\end{equation}
where $\gamma_i^{(b)}$:
\begin{equation}
\label{eq:SINR_2}
    \gamma_i^{(b)} = \frac{{\left| \mathbf{h}_b^{(i)}\mathbf{w}_{\mathbf{\Xi},b} \right|}^2}{1 + \sum_{\ell\neq b}{\left| \mathbf{h}_{b}^{(i)}\mathbf{w}_{\mathbf{\Xi},{\ell}} \right|}^2}
\end{equation}
is the SINR for the $i$-th user in the $b$-th beam. It can be noticed that, while the precoding matrix is computed assuming an equivalent channel matrix as in (\ref{eq:EquivPrecodingMatrix}), the channel matrix that is actually present between the satellite antenna feeds and the user is $\mathbf{H}^{(i)}$. The rate that can be achieved within the considered cluster is the one corresponding to the user operating with the lowest SINR among the cluster members, $\widetilde{\gamma}^{(b)}=\min_{i}\left\{\gamma_i^{(b)}\right\}$, since only with such an assumption all of the cluster members will be able to decode the received information. It is straightforward to note that the closer the SINRs among the cluster members, the better will be the performance, \emph{i.e.}, the lower will be the loss between the potential rate of a user with much higher SINR than that of the worst-case user and the actually achieved one. This closeness is directly related to that of the equivalent channel matrix $\mathbf{H}_{\mathbf{\Xi}}$, which is used to derive the precoding matrix, and the actual channel matrices $\mathbf{H}^{(i)}$, $i=1,\ldots,N_U$. Thus, it is clear that the overall system performance in terms of achievable rate strongly depends on \emph{both} the choice of the clustering algorithm and the choice of the optimisation parameter, which are the focus of this work.

In order to define an optimal clustering approach, we move from the assumption that in each beam the same number of users $N_U$ is multiplexed in the same DVB-S2 codeword during each time frame. In particular, the following assumptions hold for the users' deployment and locations: i) in the $b$-th beam, $N_G^{(b)}$ grid points are available as potential users' location, $b=1,\ldots,N_B$ (this corresponds to a spatially sampled radiation pattern, \cite{Taricco}); ii) in the generic $b$-th beam, $N_U^{(b)}$ users, $b=1,\ldots,N_B$, are assumed to be in a fixed location for a frame duration and randomly deployed according to a uniform distribution; iii) the user density within each beam is denoted by $\rho\in (0,1]$, and it is defined as the fraction of grid points to which users are allocated. Consequently, the number of users in the $b$-th beam is given by $N_U^{(b)}=\left[\rho N_G^{(b)}\right]$, $b=1,\ldots,N_B$, where $\left[\cdot\right]$ denotes the nearest integer function.

\section{Clustering Techniques for Multicast Precoding}
\label{sec:Clustering}
Clusters are defined in terms of internal homogeneity (\emph{i.e.}, similarity among cluster members) and external separation (\emph{i.e.}, dissimilarity with respect to other clusters), and both these aspects shall be properly reflected in the clustering algorithm, \cite{Book1,Book2,Book3,kmeans1,kmeansPP}. In the past, a plethora of clustering algorithm has been proposed and assessed, which can be broadly classified into one of the two following categories:
\paragraph{Partitional clustering} partitional techniques aim at defining a one-level partitioning of the initial objects into non-overlapping and non-empty clusters by means of maximisation or minimisation of a given cost function.
\begin{definition}{\textbf{Partitional Clustering}.} Given a set $\mathbf{U}=\left\{\mathbf{u}_1,\ldots,\mathbf{u}_j,\ldots,\mathbf{u}_N\right\}$ of $d$-dimensional input objects, where $\mathbf{u}_j=\left(u_{j1},\ldots,u_{jd}\right)\in\mathbb{R}^d$, with each measure $u_{ji}$ called a feature, Partitional Clustering seeks for a $K$-partition of $\mathbf{U}$, $\mathcal{C}=\left\{\mathcal{C}_1,\ldots,\mathcal{C}_K\right\}$, with $K\leq N$, such that:
\begin{enumerate}
    \item $\mathcal{C}_i\neq\emptyset$, $i=1,\ldots,K$;
    \item $\bigcup_{i=1}^K\mathcal{C}_i=\mathbf{U}$;
    \item $\mathcal{C}_i\bigcap\mathcal{C}_j=\emptyset$, $i,j=1,\ldots,K$ and $i\neq j$;
    \item the predefined cost function $\mathcal{J}$ is minimised (maximised).
\end{enumerate}
\end{definition}
\paragraph{Hierarchical clustering} these algorithms aim at identifying a tree-like partition with a single, all-inclusive cluster at the top and the single initial objects at the bottom as singleton clusters. Each intermediate level is a combination (split) of two clusters from the next lower (next higher) level in the clustering tree.
\begin{definition}{\textbf{Hierarchical Clustering}.} Given a set of $d$-dimensional input objects $\mathbf{U}=\left\{\mathbf{u}_1,\ldots,\mathbf{u}_j,\ldots,\mathbf{u}_N\right\}$, where $\mathbf{u}_j=\left(u_{j1},\ldots,u_{jd}\right)\in\mathbb{R}^d$, Hierarchical Clustering attempts to construct a tree-like, nested structure partition of $\mathbf{U}$, $\mathcal{H}=\left\{\mathcal{H}_1,\ldots,\mathcal{H}_Q\right\}$, $Q\leq N$, such that: if $C_i\in\mathcal{H}_m$, $C_j\in\mathcal{H}_{\ell}$, and $m>\ell$, then $C_i\subset C_j$ or $C_i\bigcap C_j=\emptyset$ $\forall i,j\neq i,m,\ell=1,\ldots,Q$.
\end{definition}

In the framework defined in Section~\ref{sec:ProblemStatement}, it can be noticed that Partitional Clustering algorithms provide the required user grouping criteria. The definition of the cost function $\mathcal{J}$ to be minimised (maximised) defines the homogeneity of objects within the same clusters and the separation between clusters. In the following, we consider the Sum-of-Squared-Error (SSE) cost function, which is defined as:
\begin{equation}
\label{eq:SSEcost}
\begin{split}
    \mathcal{J}\left(\mathbf{B},\mathbf{M}\right) &= \sum_{i=1}^K\sum_{j=1}^N \beta_{ij} {\left.\parallel \mathbf{u}_j-\mathbf{m}_i \parallel\right.}^2 \\
    &= \sum_{i=1}^K\sum_{j=1}^N \beta_{ij}{\left(\mathbf{u}_j-\mathbf{m}_i\right)}^T{\left(\mathbf{u}_j-\mathbf{m}_i\right)}\\
    &= \sum_{i=1}^K\sum_{j=1}^N \mathcal{J}_{ij}
\end{split}
\end{equation}
where $\mathbf{B}$ is a $K\times N$ partition matrix in which $\beta_{ij}=1$ is $1$ iff $\mathbf{u}_j\in\mathcal{C}_i$ and $0$ otherwise, $\mathcal{J}_{ij}=\beta_{ij} {\left.\parallel \mathbf{u}_i-\mathbf{m}_j \parallel\right.}^2$ is the cost function evaluated for the $i$-th object and $j$-th centroid, and $\mathbf{M}=\left(\mathbf{m}_1,\ldots,\mathbf{m}_K\right)$ is the centroid (prototype) matrix in which the $i$-th row represents the barycenter of the objects belonging to the $i$-th cluster, \emph{i.e.}:
\begin{equation}
\label{eq:centroid}
    \mathbf{m}_i = \frac{1}{N_i}\sum_{j=1}^N \beta_{ij}\mathbf{u}_j
\end{equation}
with $N_i$ being the number of objects in the $i$-th cluster. It is worthwhile noting that the partition that minimises the above cost function is optimal and leads to the \emph{minimum variance partition}, relative to the feature that has been chosen to represent the $d$-dimensional input objects.

\subsection{The k-means++ Algorithm}
One of the earliest algorithms developed within the Partitional Clustering framework is the \emph{k-means} algorithm, \cite{kmeans1}. It is based on an iterative approach that seeks a $K$-partition $\mathcal{C}$ of the input data sets by minimising the defined SSE cost function. In the proposed system architecture, we aim at identifying the clusters of users that are to be multiplexed and precoded within the same FEC codeword. Consequently, for each beam, we implement the k-means algorithm, which requires two inputs only: i) the objects to which apply the clustering procedure and the feature vector that described them, which is directly related to the metric we wish to minimise through the cost function $\mathcal{J}$; and ii) the number of clusters to be generated, $K$.\\
However, it shall be noted that, in the proposed system, for each beam $b=1,\ldots,N_B$, we have a variable number of users that depends on both the number of available grid points in the beam, $N_G^{(b)}$, and on the user density, $\rho$, yielding to $N_U^{(b)}=\rho N_G^{(b)}$ users. It can be noticed that now different beams, corresponding to input objects in the k-means algorithm, might have a quite different number of users, mainly due to the different number of grid points. Thus, an \emph{a priori} definition of a fixed number of clusters for all of the $N_B$ beams might result in a limited system flexibility and adaptability to the actual traffic requirements of each beam. To circumvent this issue, we fix the \emph{average cluster size} $S_K$, \emph{i.e.}, the average number of users per cluster within a given beam. Consequently, the number of clusters within the $b$-th beam, $K_b$, is given by the following equation:
\begin{equation}
\label{eq:UserPerCluster}
    K_b = \left\lfloor \frac{N_U^{(b)}}{S_K} \right\rfloor = \left\lfloor \frac{\left[\rho N_G^{(b)}\right]}{S_K} \right\rfloor,\ b=1,\ldots,N_B
\end{equation}
where the floor operation is required so as to have an integer number of clusters. For each beam to be served, we can now implement the k-means algorithm to identify the users grouping required by the multicast precoder. In particular, $\forall b=1,\ldots,N_B$ the algorithms first takes as input the number of required clusters for the considered beam, $K_b$, and the users feature vector $\mathbf{u}_j^{(b)}$, with $j=1,\ldots,N_U^{(b)}$, and then iterates as follows:

\begin{algorithmic}[1]
\STATE Initialise the $K_b$-partition $\mathcal{C}^{(b)}=\left\{\mathcal{C}_1^{(b)},\ldots,\mathcal{C}_{K_b}^{(b)}\right\}$ randomly or based on prior knowledge.
\STATE Compute the cluster prototype matrix
    \begin{equation*}
        \mathbf{M}^{(b)}=\left(\mathbf{m}_1^{(b)},\ldots,\mathbf{m}_{K_b}^{(b)}\right)
    \end{equation*}
\STATE Assign each object in the input data set to the nearest cluster:
\begin{equation}
\label{eq:kmeansAssign}
    \mathrm{if}\ \left.\parallel\mathbf{u}_j^{(b)}-\mathbf{m}_{\ell}^{(b)}\parallel\right.<\left.\parallel\mathbf{u}_j^{(b)}-\mathbf{m}_{i}^{(b)}\parallel\right.\Rightarrow \mathbf{u}_j^{(b)}\in\mathcal{C}_{\ell}^{(b)}
\end{equation}
with $j=1,\ldots,N_U^{(b)},\forall i,\ell=1,\ldots,K_b,i\neq\ell $.
\STATE Re-compute each cluster prototype matrix based on the current cluster members:
\begin{equation*}
    \mathbf{m}_i^{(b)} = \frac{1}{N_i}\sum_{\mathbf{u}_j^{(b)}\in\mathcal{C}_i^{(b)}} \beta_{ij}\mathbf{u}_j^{(b)},\ \forall i=1,\ldots,K
\end{equation*}
where $N_i$ is the temporary number of users in the $i$-th cluster $\mathcal{C}_i^{(b)}$.
\STATE Repeat steps 2-4 until there is no change in any of the centroids $\mathbf{m}_1^{(b)},\ldots,\mathbf{m}_{K_b}^{(b)}$.
\end{algorithmic}
Once the k-means algorithm converged and completed the users partitioning, we have, for each beam, the following $K_b$-partition:
\begin{equation}
\label{eq:Kpartition}
    \mathcal{C}^{(b)} = \bigcup_{c=1}^{K_b}\mathcal{C}_c^{(b)}, \ b=1,\ldots,N_B
\end{equation}
in which each cluster $\mathcal{C}_c^{(b)}$ is described by the set of users belonging to it and by its centroid $\mathbf{m}_c^{(b)}$. It is worthwhile noting that the partition in step 3 involves the nearest neighbour rule, in the space defined by the objects features, and it is, thus, a Voronoi tessellation. It shall be noted that, thanks to the proposed approach, the number of users for each cluster in the considered beam is not fixed anymore. As a matter of facts, we fixed the average number of users per cluster $S_K$ and then, based on the user density and spatial sampling of the beam coverage, we obtained the required number of clusters in (\ref{eq:UserPerCluster}). Based on these observations, it can be easily shown that $S_K = \sum_{c=1}^{K_b}\left|\mathcal{C}_c^{(b)}\right|$.

The performance of the k-means algorithm strongly depends on the choice of the initial partition centroids and this is also related to two of its major shortcomings: i) the worst case running time of the algorithm is super-polynomial in the input size; and ii) the clustering result can be arbitrarily bad with respect to the objective function compared to the optimal clustering. Several solutions have been proposed in the literature to improve the performance based on the initial selection of centroids. For instance, in Forgy's method the initial partition is generated by randomly selecting $K$ points as centroids and then separating the remaining points based on their distance from these seeds, \cite{kmeans2}. In MacQueen's approach, the first centroid is the most centrally located and has the smallest vector sum of distances to other points (\emph{i.e.}, it is the barycenter of the input set $\mathbf{X}$); then, the point that achieves the maximum decrease of the cost function is selected as the second centroid and the procedure is iterated until $K$ initial centroids are selected, \cite{kmeans3}.\\
In 2007, D. Arthur and S. Vassilvitskii defined an approximation algorithm for the NP-hard k-means problem, which then became known as the \emph{k-means++ algorithm}, \cite{kmeansPP}. In this algorithm, steps 2-5 are the same as in the traditional k-means, but the initial centroid selection is based on a probabilistic approach. In particular, when initialising the $b$-th beam in the considered system, the following algorithm is implemented:
\begin{algorithmic}[1]
\STATE The first centroid $\mathbf{m}_1^{(b)}$ is randomly chosen among the users.
\STATE For each data object $\mathbf{u}_j^{(b)}$, $j=1,\ldots,N_U^{(b)}$, compute the cost function between it and each of the centroids that have been defined until now:
    \begin{equation*}
        \mathcal{J}_{ij} = \mathcal{J}\left(\mathbf{u}_j^{(b)},\mathbf{m}_i^{(b)}\right)
    \end{equation*}
$\forall j=1,\ldots,N_U^{(b)}, \forall i=1,\ldots,N_{all}$ where $N_{all}< K_b$ is the number of previously identified centroids, and with $\mathcal{J}_{ij}$ that is defined in (\ref{eq:SSEcost}) and it depends on the metric to be minimised that has been chosen.
\STATE Select the minimum distance from step 2 for each object:
\begin{equation*}
    \mathcal{J}\left(\mathbf{u}_j^{(b)}\right) = \min_{i\in[1,N_{all}]} \mathcal{J}\left(\mathbf{u}_j^{(b)},\mathbf{m}_i^{(b)}\right),\ j=1,\ldots,N
\end{equation*}
\STATE Randomly choose the next centroid with a probability proportional to $\mathcal{J}\left(\mathbf{u}_j^{(b)}\right)$.
\STATE Iterate steps 2-4 until the pre-defined number of centroids, $K_b$, is reached.
\end{algorithmic}
Numerical simulations show that the k-means++ algorithm can perform twice as fast as the traditional k-means and also provide significantly better clustering solutions, \cite{kmeansPP}.

The clustering operation is performed at the system GW once all the feature vectors of each user in the $N_B$ beams is known. Two different metrics, \emph{i.e.}, object features that drive the minimisation of the cost function $\mathcal{J}$, will be considered and described in the following.
\subsubsection{Euclidean Distance}
Assuming that the GW knows the location of each user in each beam, the clustering algorithm can exploit this knowledge so as to group together users based on the Euclidean distance. In particular, let us denote by $\mathbf{v}_{b,j}\in\mathbb{R}^2$ the location of the $j$-th user in the $b$-th beam, with $b=1,\ldots,N_B$ and $i=1,\ldots,N_U^{(b)}$. By considering these bi-dimensional location vectors as input features in the k-means++ algorithm, \emph{i.e.}, $\mathbf{u}_j^{(b)}=\mathbf{v}_{b,j}$ in (\ref{eq:kmeansAssign}), we obtain a $K_b$-partition of the users in the generic $b$-th beam, $\mathcal{C}^{(b)}=\left\{\mathcal{C}_1^{(b)},\ldots,\mathcal{C}_{K_b}^{(b)}\right\}$ such that:
\begin{equation}
\label{eq:DistPartition}
    \left.\parallel\mathbf{v}_{b,j}-\mathbf{m}_{\ell}^{(b)}\parallel\right.<\left.\parallel\mathbf{v}_{b,j}-\mathbf{m}_{i}^{(b)}\parallel\right.,\ \forall i\neq\ell \Rightarrow \mathbf{v}_{b,j}\in\mathcal{C}_{\ell}^{(b)}
\end{equation}
where $\mathbf{m}_i^{(b)}$, $i=1,\ldots,K_b$ denote the locations of the $K_b$-partition centroids for the $b$-th beam. It is straightforward to note that when the k-means++ algorithm converges and provides the beam clusters, the $K_b$ centroids correspond to the barycenter of each cluster.

\subsubsection{Channel Coefficients}
A different approach for the user clustering is based on the observation that the rate assigned to the users of the generic $c$-th cluster $\mathcal{C}_c^{(b)}$ in the $b$-th beam, with $b=1,\ldots,N_B$ and $c=1,\ldots,K_b$, is constrained by the user with the lowest SINR, since all of them are precoded in the same codeword and frame, \emph{i.e.}, they must use the FEC codeword length and Modulation and Coding scheme (ModCod) that allows the user in the worst channel conditions to decode its information. Let us denote the minimum SINR among the cluster users by $\widetilde{\gamma}_c^{(b)}$, for $b=1,\ldots,N_B$ and $c=1,\ldots,K_b$:
\begin{equation}
\label{eq:minSINRuser}
    \widetilde{\gamma}_c^{(b)} = \min_{m\in\mathcal{C}_c^{(b)}}\left\{ \gamma_m^{(b)} \right\},\ m = 1,\ldots,\left|\mathcal{C}_c^{(b)}\right|
\end{equation}
where $\gamma_m^{(b)}$, denoting the SINR of the $m$-th user in the $b$-th beam, is defined in (\ref{eq:SINR_2}) and it provides the FEC codeword length and ModCod, \emph{i.e.}, the rate $\eta_c^{(b)}$, that will be assigned to the users belonging to $\mathcal{C}_c^{(b)}$.\\
Since the SINR value of each user in the cluster is defined by the precoding matrix, and the precoding matrix is built by averaging on the channel coefficients of the cluster members as shown in (\ref{eq:EquivChannelMatrix})-(\ref{eq:EquivPrecodingMatrix}), we can implement the k-means++ algorithm so as to account for this optimisation. In particular, each user can be described by a $N_B$-dimensional feature vector given by its channel coefficients, $\mathbf{h}_b^{(j)}$, with $j=1,\ldots,N_U^{(b)}$, where we dropped the dependency on the user location $\mathbf{v}_{b,j}$ for the sake of simplicity. Before implementing the k-means++ clustering, it shall be noted that the precoding matrix computed through the pragmatic approach proposed in \cite{Taricco} is obtained by means of a column-wise normalisation of the equivalent channel matrix $\mathbf{H}_{\mathbf{\Xi}}$. Thus, in order to correctly relate the clustering metric to the effect of the precoding matrix on the received signals, we normalise the feature vectors to be considered for each user as well. In particular, for the generic $j$-th user, $j=1,\ldots,N_U^{(b)}$ in the $b$-th beam, $b=1,\ldots,N_B$, the following normalised vector is considered as input feature for the k-means++ clusterisation:
\begin{equation}
\label{eq:ChannelFeature}
\begin{split}
    \mathbf{u}_j^{(b)}=\hat{\mathbf{h}}_b^{(j)} &= \left(\hat{h}_{1,b}^{(j)},\ldots,\hat{h}_{N_B,b}^{(j)}\right) \\
    \hat{h}_{k,b}^{(j)} &= \frac{h_{k,b}^{(j)}}{\sqrt{\sum_{k=1}^{N_B}{\left| h_{k,b}^{(j)} \right|}^2}}
\end{split}
\end{equation}
The k-means++ algorithm is then implemented by minimising the same cost function as for the previous approach, \emph{i.e.}, the Euclidean distance, which this time is computed in the $N_B$-dimensional space defined by the above normalised channel coefficients. Thus, from (\ref{eq:kmeansAssign}), we obtain the following rule:
\begin{equation}
\label{eq:ChannPartition}
    \left.\parallel\hat{\mathbf{h}}_b^{(j)}-\mathbf{m}_{\ell}^{(b)}\parallel\right.<\left.\parallel\hat{\mathbf{h}}_b^{(j)}-\mathbf{m}_{i}^{(b)}\parallel\right.,\ \forall i\neq\ell \Rightarrow \mathbf{v}_{b,j}\in\mathcal{C}_{\ell}^{(b)}
\end{equation}
where, differently from (\ref{eq:DistPartition}), $\mathbf{m}_i^{(b)}$, $i=1,\ldots,K_b$, denotes the $b$-th beam cluster centroids in the $N_B$-dimensional space defined by the normalised channel coefficients and the statement $\mathbf{v}_{b,j}\in\mathcal{C}_{\ell}^{(b)}$ is identical to that in (\ref{eq:DistPartition}) to unambiguously define that the $j$-th user is assigned to the $\ell$-th cluster.

\begin{figure}[!t]
\centering
\includegraphics[width=0.45\columnwidth]{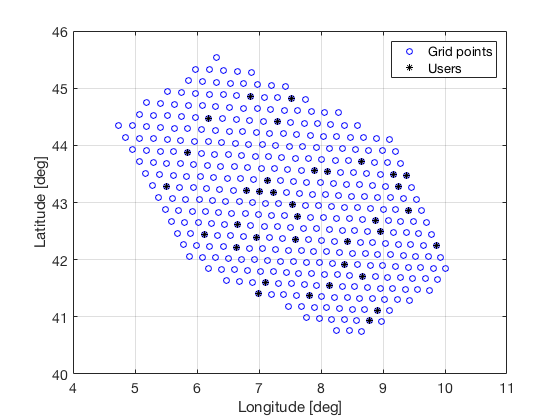}
\caption{Example of beam spatial sampling and users deployment. Setup: $\rho = 0.1$.}
\label{fig:Users}
\end{figure}

\begin{figure}[!ht]
     \subfloat[Euclidean distance method.\label{fig:beam30_dist}]{%
       \includegraphics[width=0.45\textwidth]{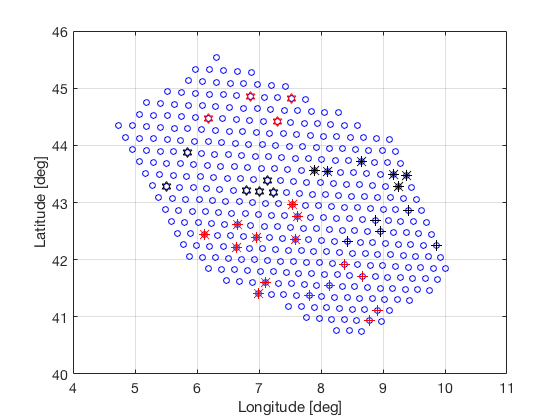}
     }
     \hfill
     \subfloat[Channel coefficients method.\label{fig:beam30_chann}]{%
       \includegraphics[width=0.45\textwidth]{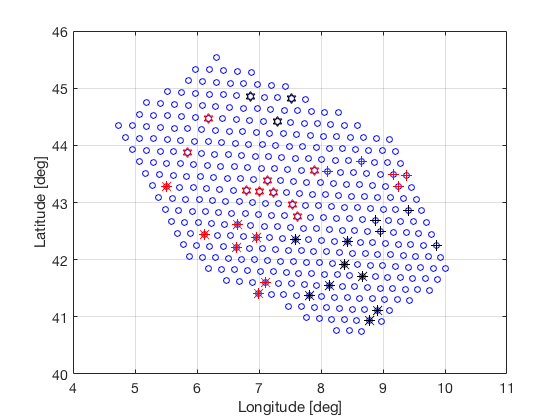}
     }
     \caption{Example of clustering with the Euclidean and Channel methods. Setup: $\rho=0.1$, $S_K=6$.}
     \label{fig:dummy}
\end{figure}

Figure~\ref{fig:Users} shows an example of randomly deployed users in the spatially sampled locations for beam $b=30$ and with user density $\rho=0.1$. Based on this deployment, Figures~\ref{fig:beam30_dist}-\ref{fig:beam30_chann} show the clustering outcome when considering an average cluster size $S_K=6$ with the Euclidean and Channel methodologies. The difference in the clusters composition between the two methods is clear and this already provides a critical information: users that are closer in terms of their location, \emph{i.e.}, users clustered together with the Euclidean method, not necessarily show the same closeness in terms of channel coefficients.

%

\subsection{System scheduling}
\begin{figure}[!t]
\centering
\includegraphics[width=0.5\columnwidth]{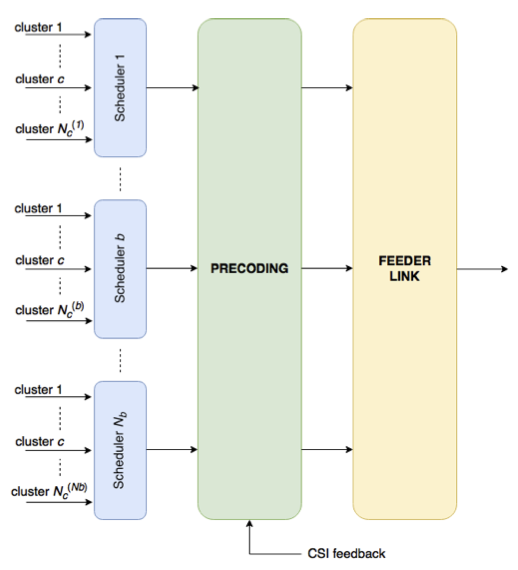}
\caption{Cluster scheduling.}
\label{fig:ClusterScheduler}
\end{figure}

After the implementation of the k-means++ algorithm with any of the two metrics defined above, the system GW shall simultaneously serve one cluster for each beam in a single time frame. As shown in Fig.~\ref{fig:ClusterScheduler}, we thus have a cluster scheduler for each of the beam to be served, which provides the index $c$ corresponding to the cluster to be served in each of the $N_B$ beams. Since we have a variable number of clusters per beam, $K_b$ with $b=1,\ldots,N_B$, there will be beams for which all of the clusters have been already served while others still have unserved groups of users. In that case, the scheduler of the beams that already have completed a round-robin of their cluster randomly select a cluster for additional frames.

\begin{figure}[!t]
\centering
\includegraphics[width=0.45\columnwidth]{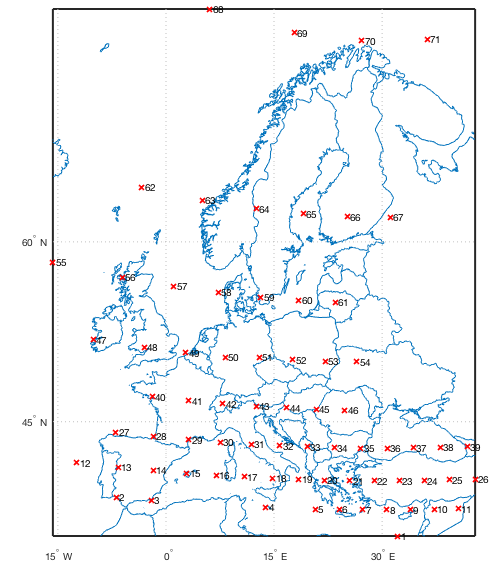}
\caption{Multi-beam satellite system covering Europe through $71$ beams. The red crosses identify the beam centers.}
\label{fig:multi-beamPlot}
\end{figure}

\begin{figure}[!t]
\centering
\includegraphics[width=0.5\columnwidth]{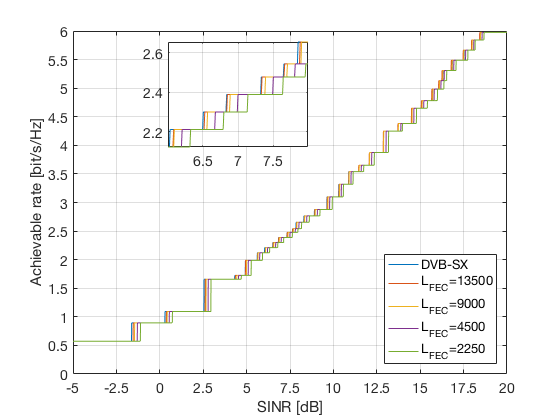}
\caption{Modulation and Coding schemes (ModCod) used to derive the rate as a function of the SINR in (\ref{eq:NumericalRate}).}
\label{fig:MdoCod}
\end{figure}

\section{Numerical Results}
\label{sec:Results}
In this section, we compare the performance of the two clustering approaches defined in the previous section, which will be referred to as \emph{Euclidean} and \emph{Channel} in the following, in terms of the average rate with which the system is able to serve its beams. This will provide a significant insight on the parameters that drive the system performance, which can lead to several degrees of optimisation of the system design.\\
We consider a multi-beam satellite system covering the whole of Europe through $N_B=71$ beams, as shown in Fig.~\ref{fig:multi-beamPlot}, and the simulation parameters are listed in Table~\ref{tab:SimulationParam}. During each Monte Carlo iterations, $\rho N_G^{(b)}$ users are randomly dropped in each beam with a uniform distribution, as shown in Fig.~\ref{fig:Users}. By implementing the k-means++ algorithm previously described, with both metrics, we obtain a $K_b$-partition of each beam, $b=1,\ldots,N_B$, \emph{i.e.}, for each beam we have $K_b$ clusters to be served in $K_b$ successive time frames, together with those of the other beams. Since, as previously highlighted, different beams will have a different number of clusters, when a certain beam already served all of its users it randomly picks an already served cluster to serve it again. By denoting the maximum number of clusters throughout the $71$ beams as $K=\max_b\{K_b\}$, this means that we simulate a total of $400\cdot K$ frames in which $N_B$ clusters with variable size are served together. The input parameters for the numerical simulations are:
\begin{itemize}
    \item the average number of users per cluster, $S_K$;
    \item the user density $\rho$;
    \item the FEC codeword length, $L_{FEC}$.
\end{itemize}
For the generic $c$-th cluster in the $b$-th beam and $n$-th MC iteration, we obtain a rate value that depends on the above paramters, $\eta_{c,n}^{(b)}(S_K,\rho,L_{FEC})$, which is a function of the minimum SINR among the cluster members, \emph{i.e.}:
\begin{equation}
\label{eq:NumericalRate}
    \eta_{c,n}^{(b)}(S_K,\rho,L_{FEC}) = f\left(\widetilde{\gamma}_c^{(b)}\right)
\end{equation}
The function $f$ reflects the different ModCods that are considered and they are shown in Fig.~\ref{fig:MdoCod}. As discussed in the previous sections, the above minimum SINR depends on the clustering metric used to describe the users feature space, the average cluster size, and the user density, since these are the design parameters that affect the performance of the k-means++ algorithm. In the following, we provide a thorough insight on the impact that each of these parameters has on the overall system performance, \emph{i.e.}, the average rate. This value is obtained by averaging over MC iterations, simulated clusters, and number of beams:
\begin{equation}
\label{eq:AverageRate}
    \overline{\eta}(S_K,\rho,L_{FEC}) = \mathbb{E}_{n,c,b}\left\{\eta_{c,n}^{(b)}(S_K,\rho,L_{FEC})\right\}
\end{equation}
In the following, we analyse the numerical simulations in terms of the above average with and without precoding for both the Euclidean and the Channel methodologies as a function of the average cluster size. Moreover, we also provide the average rate in (\ref{eq:AverageRate}) as a function of the average distance from the cluster centroids and, finally, compare the performance of the two proposed clusterisation approaches.

\begin{table}[t]
\renewcommand{\arraystretch}{1.3}
\caption{Numerical simulation parameters}
\label{tab:SimulationParam}
\centering
\begin{tabular}{|c|c|}
\hline
\bfseries Parameter & \bfseries Value\\
\hline\hline
Carrier frequency & $19.5$ GHz\\
\hline
Receiving antenna diameter & $0.6$ m\\
\hline
Receiving antenna efficiency & $0.6$\\
\hline
Antenna losses & $2.55$ dB\\
\hline
GEO satellite longitude & $30^{\circ}$\\
\hline
Satellite transmitted power & $P_{sat}=90$W\\
\hline
Dual polarisation power & $P_{tx}=P_{sat}/2=45$ W\\
\hline
$N_B$ & $71$\\
\hline
$\rho$ & $0.1$, $0.2$, $0.4$, $0.6$, $0.8$, $1.0$\\
\hline
$L_{FEC}$ & $2250$, $4500$, $9000$, $13500$\\
\hline
$S_K$ & $1$ (unicast), $2,4,\ldots,16$\\
\hline
Target Bit Error Probability & $10^{-4}$\\
\hline
\end{tabular}
\end{table}

\subsection{Average Rate vs. Average Cluster Size}
\begin{figure}[!ht]
     \subfloat[Euclidean distance method.\label{fig:Rate_Dist_10}]{%
       \includegraphics[width=0.45\textwidth]{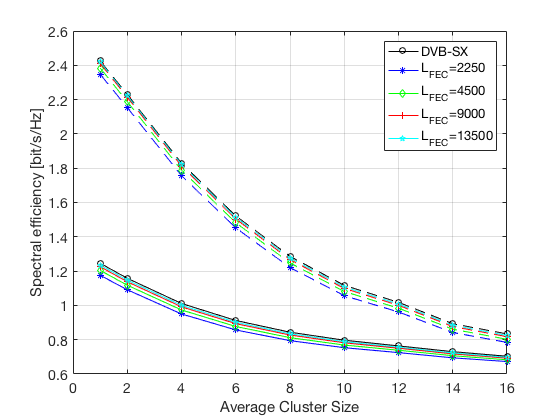}
     }
     \hfill
     \subfloat[Channel coefficients method.\label{fig:Rate_Ch_10}]{%
       \includegraphics[width=0.45\textwidth]{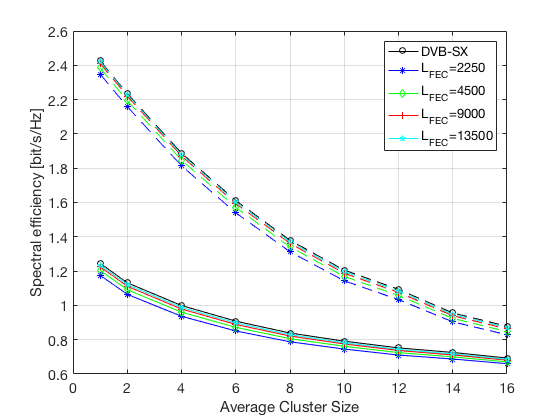}
     }
     \caption{Average rate per beam with (dashed line) and without (solid line) precoding. Setup: $\rho = 0.1$.}
     \label{fig:Rate_10}
\end{figure}

\begin{figure}[!ht]
     \subfloat[Euclidean distance method.\label{fig:Rate_Dist_60}]{%
       \includegraphics[width=0.45\textwidth]{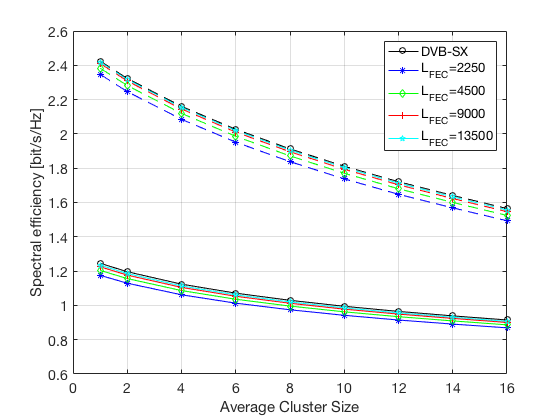}
     }
     \hfill
     \subfloat[Channel coefficients method.\label{fig:Rate_Ch_60}]{%
       \includegraphics[width=0.45\textwidth]{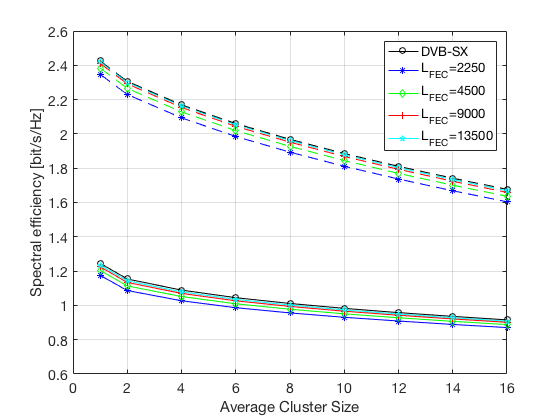}
     }
     \caption{Average rate per beam with (dashed line) and without (solid line) precoding. Setup: $\rho = 0.6$.}
     \label{fig:Rate_60}
\end{figure}

\begin{figure}[!ht]
     \subfloat[Euclidean distance method.\label{fig:Rate_Dist_100}]{%
       \includegraphics[width=0.45\textwidth]{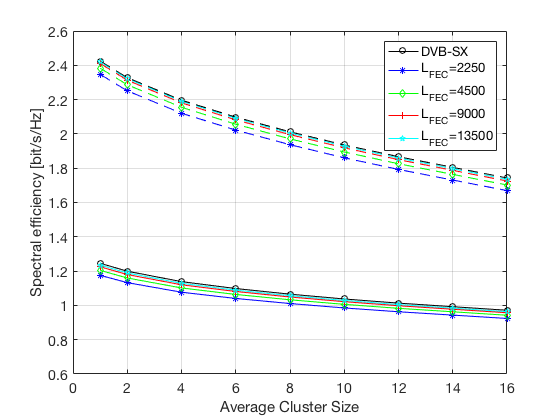}
     }
     \hfill
     \subfloat[Channel coefficients method.\label{fig:Rate_Ch_100}]{%
       \includegraphics[width=0.45\textwidth]{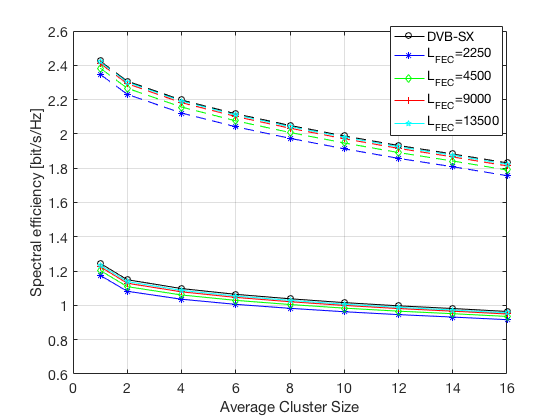}
     }
     \caption{Average rate per beam with (dashed line) and without (solid line) precoding. Setup: $\rho = 1.0$.}
     \label{fig:Rate_100}
\end{figure}

Figures~\ref{fig:Rate_10}-\ref{fig:Rate_100} provide the average rate in each beam, as defined in (\ref{eq:AverageRate}), as a function of the average cluster size $S_K$, both with and without precoding and for different ModCods. The following behaviours can be observed:
\begin{itemize}
    \item the average rate decreases for increasing values of the average cluster size. This is due to the fact that, increasing $S_K$, the clusters tend to have larger sizes and, thus, more users with different channel coefficients are grouped into the same FEC codeword. Since the user with the lowest SINR is the one driving the serving rate, as outlined in (\ref{eq:minSINRuser}), the performance tends to be worse. The relation of this behaviour to the difference in the channel coefficients of the users clustered together is substantiated by observing that, with larger values of the user density $\rho$ (see Figures \ref{fig:Rate_Dist_100}-\ref{fig:Rate_Ch_100} for $\rho=1.0$), the loss for increasing $S_K$ is lower: in this case, in fact, the clusterisation algorithm can find users that are closer one to each other with both the Euclidean and the Channel methodologies.
    \item The average rate is significantly larger when precoding is implemented, confirming that significant performance benefits can be obtained when implementing this technique. This can be noticed in Tables~\ref{tab:PrecodingGain_EU}-\ref{tab:PrecodingGain_CH}, which provide the gains obtained with respect to the absence of precoding for different values of the $\rho$ and $S_K$ for DVB-SX ModCods.
    \item The average rate tends to decrease also when not implementing precoding for increasing values of the average cluster size. Although this behaviour might seem puzzling, it shall be noted that, even without precoding, the users are still clustered together and, thus, also in this case there is a loss related to the user with lower SINR.
    \item It can be noticed that, in general, the Channel approach provides a better performance with a gain that is larger for lower values of the user density. This can be explained by observing that, in the Euclidean approach, a low user density implies that clusters are built with users that are located far away from each other, which results in significantly different channel coefficients and different SINR values. When increasing the average cluster size with low user densities, the difference between the two approaches decreases, since even in the channel coefficients space it becomes more difficult to find similar users (the number of users is limited and spread in a wide area)
\end{itemize}

\begin{table*}[t]
\renewcommand{\arraystretch}{1.3}
\caption{Precoding gain: Euclidean}
\label{tab:PrecodingGain_EU}
\centering
\begin{tabular}{|c|c|c|c|c|c|c|c|c|c|}
\hline
$\rho$ & $S_K=1$ & $S_K=2$ & $S_K=4$ & $S_K=6$ & $S_K=8$ & $S_K=10$ & $S_K=12$ & $S_K=14$ & $S_K=16$\\
\hline\hline
$0.1$ & $94.71\%$ & $92.55\%$ & $80.98\%$ & $66.89\%$ & $52.38\%$ & $40.14\%$ & $33.27\%$ & $22.30\%$ & $18.48\%$\\
\hline
$0.2$ & $94.62\%$ & $93.58\%$ & $87.59\%$ & $79.52\%$ & $70.92\%$ & $62.30\%$ & $55.03\%$ & $47.70\%$ & $39.63\%$\\
\hline 
$0.4$ & $94.69\%$ & $94.11\%$ & $91.07\%$ & $86.80\%$ & $81.81\%$ & $76.55\%$ & $71.81\%$ & $66.73\%$ & $62.18\%$ \\
\hline
$0.6$ & $94.67\%$ & $94.21\%$ & $92.27\%$ & $89.22\%$ & $85.75\%$ & $82.10\%$ & $78.37\%$ & $74.77\%$ & $71.02\%$ \\
\hline
$0.8$ & $94.66\%$ & $94.17\%$ & $92.75\%$ & $90.40\%$ & $87.76\%$ & $85.00\%$ & $82.15\%$ & $79.13\%$ & $76.23\%$\\
\hline
$1.0$ & $94.66\%$ & $94.10\%$ & $92.94\%$ & $91.01\%$ & $88.87\%$ & $86.58\%$ & $84.25\%$ & $81.89\%$ & $79.37\%$ \\
\hline
\end{tabular}
\end{table*}

\begin{table*}[t]
\renewcommand{\arraystretch}{1.3}
\caption{Precoding gain: Channel}
\label{tab:PrecodingGain_CH}
\centering
\begin{tabular}{|c|c|c|c|c|c|c|c|c|c|}
\hline
$\rho$ & $S_K=1$ & $S_K=2$ & $S_K=4$ & $S_K=6$ & $S_K=8$ & $S_K=10$ & $S_K=12$ & $S_K=14$ & $S_K=16$\\
\hline\hline
$0.1$ & $94.75\%$ & $97.46\%$ & $88.94\%$ & $77.41\%$ & $64.47\%$ & $52.33\%$ & $45.32\%$ & $31.74\%$ & $26.73\%$\\
\hline
$0.2$ & $94.66\%$ & $98.59\%$ & $95.02\%$ &	$88.86\%$ & $81.74\%$ & $74.20\%$ & $68.16\%$ & $61.61\%$ & $53.68\%$\\
\hline
$0.4$ & $94.68\%$ & $99.45\%$ & $98.17\%$ & $95.12\%$ & $91.43\%$ & $87.35\%$ & $83.45\%$ & $78.94\%$ & $74.85\%$\\
\hline
$0.6$ & $94,72\%$ & $99.90\%$ & $99.19\%$ & $97.16\%$ & $94.67\%$ & $91.88\%$ & $89.02\%$ & $86.08\%$ & $83.13\%$\\
\hline
$0.8$ & $94.71\%$ & $100.31\%$ & $99.74\%$ & $98.18\%$ & $96.31\%$ & $94.19\%$ & $92.00\%$ & $89.61\%$ & $87.47\%$ \\
\hline
$1.0$ & $94.69\%$ & $100.72\%$ & $100.14\%$ & $98.90\%$ & $97.34\%$ & $95.66\%$ & $93.73\%$ & $91.80\%$ & $89.78\%$\\
\hline
\end{tabular}
\end{table*}

%

%

%

\begin{figure}[!ht]
     \subfloat[Euclidean distance method.\label{fig:DistRate_Eu_0}]{%
       \includegraphics[width=0.45\textwidth]{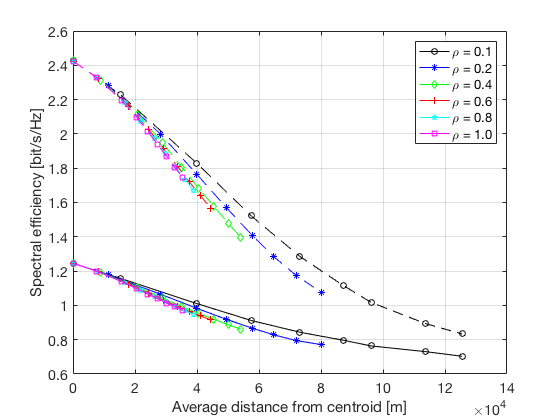}
     }
     \hfill
     \subfloat[Channel coefficients method.\label{fig:DistRate_Ch_0}]{%
       \includegraphics[width=0.45\textwidth]{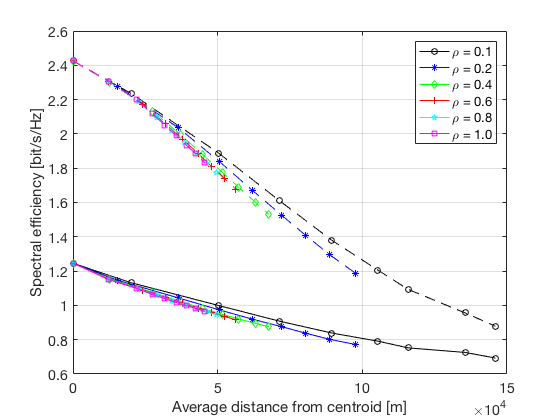}
     }
     \caption{Average rate per beam as a function of the average distance from the cluster centroid with (dashed line) and without (solid line) precoding. Setup: DVB-SX ModCods.}
     \label{fig:DistRate_0}
\end{figure}

%

\begin{figure}[!ht]
     \subfloat[Euclidean distance method.\label{fig:DistRate_Eu_2250}]{%
       \includegraphics[width=0.45\textwidth]{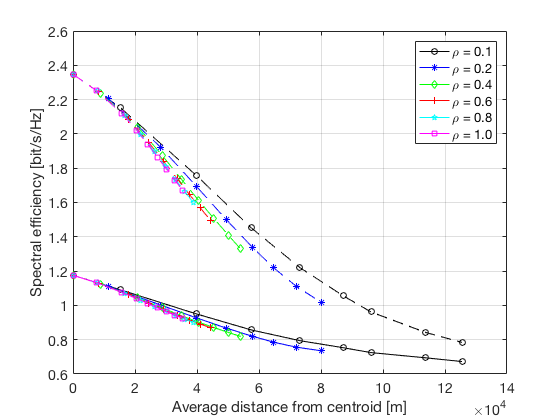}
     }
     \hfill
     \subfloat[Channel coefficients method.\label{fig:DistRate_Ch_2250}]{%
       \includegraphics[width=0.45\textwidth]{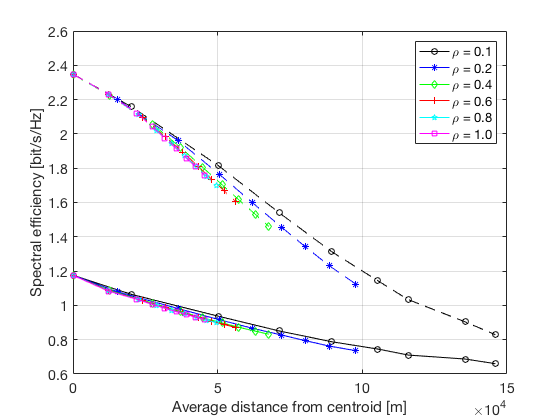}
     }
     \caption{Average rate per beam as a function of the average distance from the cluster centroid with (dashed line) and without (solid line) precoding. Setup: $L_{FEC}=2250$.}
     \label{fig:DistRate_2250}
\end{figure}

%

\begin{figure}[!ht]
     \subfloat[Euclidean distance method.\label{fig:DistRate_Eu_13500}]{%
       \includegraphics[width=0.45\textwidth]{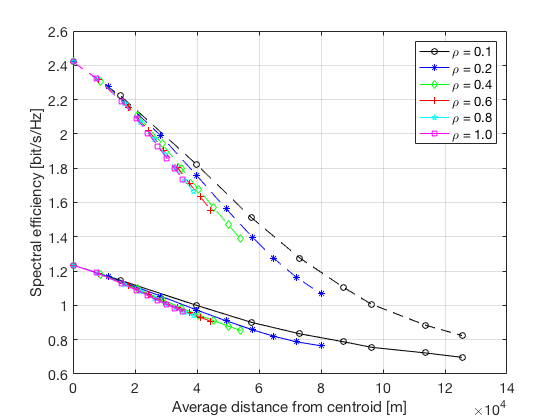}
     }
     \hfill
     \subfloat[Channel coefficients method.\label{fig:DistRate_Ch_13500}]{%
       \includegraphics[width=0.45\textwidth]{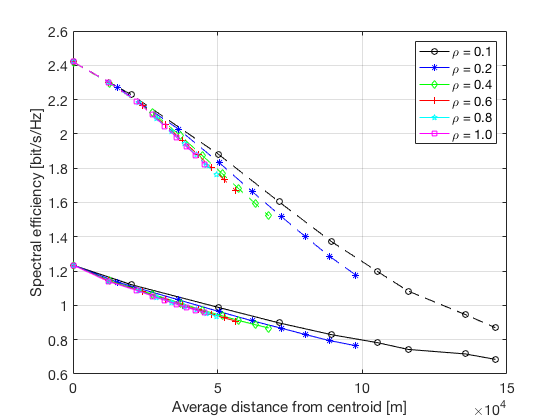}
     }
     \caption{Average rate per beam as a function of the average distance from the cluster centroid with (dashed line) and without (solid line) precoding. Setup: $L_{FEC}=13500$.}
     \label{fig:DistRate_13500}
\end{figure}

\subsection{Average Rate vs. Average Centroid Distance}
\begin{figure}[!t]
\centering
\includegraphics[width=0.5\columnwidth]{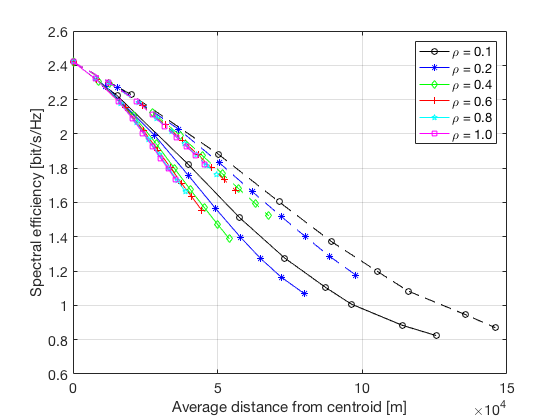}
\caption{Average rate per beam as a function of the average distance from the cluster centroid with the Euclidean method (solid line) and the channel coefficients method (dashed line). Setup: with precoding, $L_{FEC}=13500$.}
\label{fig:DistRate_Comp_13500}
\end{figure}

We now compare the average rate per beam, with different ModCods and both clustering solutions, is provided as a function of the average distance between the cluster users and its corresponding centroid. Considering the $c$-th cluster of the $b$-th beam, the distance of the generic $j$-th user of the cluster from its centroid can be computed based on (\ref{eq:SSEcost}) as follows:
\begin{equation}
    d_{cj}=\ {\left.\parallel \mathbf{v}_j^{(b)}-\mathbf{m}_c^{(b)} \parallel\right.},\ j\in \mathcal{C}_c^{(b)}
\end{equation}
The average distance is thus computed by averaging over the MC iterations, beams, clusters, and users:
\begin{equation}
    \overline{D}\left(\rho\right) = \mathbb{E}_{b,c,j}\left\{d_{cj}\right\}
\end{equation}
where we highlighted that the average distance of the cluster users from their centroid is strictly dependent on the user deployment density.\\
From Figures~\ref{fig:DistRate_0}-\ref{fig:DistRate_13500}, the following observations can be made:
\begin{itemize}
    \item a given average rate per beam is achieved with a lower average distance between the cluster users and their centroids when considering larger user densities.
    \item For a fixed average distance between the cluster users and their centroids, larger average rate values are obtained with lower user densities. 
\end{itemize}
The two above behaviours can be explained as follows. As outlined in the previous subsection, the overall system performance is driven by the number of users that are grouped together in the same FEC codeword. When increasing the user density in each beam, large clusters are built even when considering a limited average distance between them and the cluster centroid, while larger distances are required to reach the same clusters cardinality with lower user densities.\\
Furthermore, it can also be observed that, with increasing user densities, the behaviour of the average rate for increasing distances tends to be the same. As a matter of facts, when almost all of the spatially sampled locations in each beam are available for user deployment, the same average cluster size is achieved for a fixed average distance between users and centroids and, thus, the same serving rate can be achieved.\\
Finally, in Figure~\ref{fig:DistRate_Comp_13500}, the average rate per beam is observed as a function of the average distance between the cluster users and the corresponding centroid. It can be noticed that, for a specific average rate, the Euclidean methodology provides users that are significantly closer to each other, while with the Channel approach users are more distant from their centroids. This is due to the fact that, in the latter methodology, users are grouped together based on their distance in the channel coefficients space, which is not only affected by the geographical distance but also by the antenna radiation patterns. Consequently, in order to find users that can be grouped together, the Channel methodology requires to search users in a wider area relative to the cluster centroid.

%
%

\section{Conclusions}
\label{sec:Conclusion}
In this paper, we focused on the users grouping in the context of multicast precoding algorithms for multi-beam satellite systems. In particular, moving from early works, we defined a mathematical framework for the design of multicast precoding systems based on clustering theory and algorithms. In the proposed system, and differently from previous studies, a variable number of users per beam is assumed, as well as a variable number of users per group to be precoded within the same frame. Efficient clustering solutions based on the well-known k-means algorithm have been implemented aimed at providing minimum variance partitions of the initial users pool. In particular, two metrics have been considered: i) the Euclidean distance, \emph{i.e.}, \emph{Euclidean} method; and ii) the distance in the multi-dimensional space of user channel coefficients, \emph{i.e.}, \emph{Channel} method. Numerical simulations showed that the achievable rate provided by the multicast precoding system when implementing the Channel methodology is actually better with respect to that obtained with the Euclidean approach, with different codeword lengths, user densities, and average cluster sizes. However, both approaches significantly improve the performance with respect to a system in which no precoding is implemented. In particular, these gains range from approximately $20\%$ up to $100\%$, depending on the system configuration. In addition, we also showed the dependency of the average rate obtained with the two methodologies as a function of the average distance between the cluster users and their centroid. In this case, it has been highlighted that the average distance when implementing the Channel method is significantly greater than that obtained with the Euclidean approach, for a fixed rate. This behaviour indicates that users that are close in the Euclidean space are not necessarily close in the channel coefficients multi-dimensional space. Since the latter is the space driving the overall achievable throughput, the Channel methodology results in a better clustering from the data rate perspective.


\section*{Acknowledgment}
This work has been partially supported by European Space Agency (ESA) funded project OPTIMUS (``OPtimized transmission TechnIques for satcoM UnicaSt Interactive traffic''), contract 4000116421/15/NL/FE. The views of the authors of this paper do not reflect the views of ESA.

\ifCLASSOPTIONcaptionsoff
  \newpage
\fi

\end{document}